TITLE

# A network-based approach for surveillance of occupational health exposures

AUTHORS AND FULL ADDRESSES

**Laurie Faisandier**
Laboratoire Environnement et Prédiction de la Santé des Populations-TIMC, Université Joseph Fourier, Grenoble, France
E-mail LFaisandier@chu-grenoble.fr

**Vincent Bonneterre**, Laboratoire Environnement et Prédiction de la Santé des Populations-TIMC, Université Joseph Fourier, and Service de Médecine et Santé au Travail, CHU Grenoble, France

**Régis de Gaudemaris**, Laboratoire Environnement et Prédiction de la Santé des Populations-TIMC, Université Joseph Fourier, and Service de Médecine et Santé au Travail, CHU Grenoble, France

**Dominique J. Bicout**, Biomathématiques et Epidémiologie, Laboratoire Environnement et Prédiction de la Santé des Populations-TIMC, UMR CNRS 5525 Université Joseph Fourier, Ecole Nationale Véterinaire de Lyon, 69280 Marcy l'Etoile, France
E-mail Dominique.Bicout@imag.fr

**RNV3P**\*, C. Doutrellot-Philippon (Amiens), D. Penneau-Fontbonne Y. Roquelaure (Angers), I. Tahon (Besançon), P. Brochard, C. Verdun-Esquer (Bordeaux), J.D. Dewitte (Brest), M. Letourneux (Caen), M.F. Marquignon (Cherbourg), A. Chamoux, L. Fontana (Clermont-Ferrand), J.C. Pairon (Créteil), H.J. Smolik (Dijon), J. Ameille, A. d'Escatha (Garches), A. Maitre, E. Michel (Grenoble), A. Gislard (Le Havre), P. Frimat, C. Nisse (Lille), D. Dumont (Limoges), A. Bergeret, J.C. Normand (Lyon), M.P. Le Hucher-Michel (Marseille), C. Paris (Nancy), D. Dupas, C. Geraut (Nantes), D. Choudat (Paris – Cochin), R. Garnier (Paris – Fernand Widal), D. Leger (Paris – Hotel-Dieu), E. Ben-Brik (Poitiers), F. Deschamps (Reims), A. Caubet, C. Verger (Rennes), J.F. Caillard, J.G. Gehanno (Rouen), D. Faucon (Saint-Etienne), A. Cantineau, (Strasbourg), J.M. Soulat (Toulouse), G. Lasfargues (Tours)

\* RNV3P is indexed in PubMed

SUMMARY

In the context of surveillance of health problems, the research carried out by the French national occupational disease surveillance and prevention network (Réseau National de Vigilance et de Prévention des Pathologies Professionnelles, RNV3P) aims to develop, among other approaches, methods of surveillance, statistical analysis and modeling in order to study the structure and change over time of relationships between disease and exposure, and to detect emerging disease-exposure associations. In this perspective, this paper aims to present the concept of the « exposome » and to explain on what bases it is constructed. The exposome is defined as a network of relationships between occupational health problems that have in common one or several elements of occupational exposure (exposures, occupation and/or activity sector). The paper also aims to outline its potential for the study and programmed surveillance of composite disease-occupational exposure associations. We illustrate this approach by applying it to a sample from the RNV3P data, taking malignant tumours and focusing on the subgroup of non-Hodgkin lymphomas.

KEY INDEX WORDS

Composite exposures; network; exposome; occupational health

MAIN BODY OF THE WORK

1. INTRODUCTION

Surveillance of work-related diseases and exposures is a major issue of public health, in particular in order to identify and prevent new risks. In occupational health, surveillance conventionally involves epidemiological follow-up of indicators in the general population (accidents and occupational diseases) and specific cohort or case-control studies, if numbers are sufficient for the expected statistical power. In this field of study of occupational exposure, health surveillance is confronted by several methodological challenges: how to take into account all the components of exposure (single or multiple occupational exposures, occupations, activity sectors), how to cover diseases where the number of cases may be too few to be approached by conventional epidemiological methods, and how to pinpoint potentially emerging composite disease-exposure associations.

In such a context, monitoring and epidemiological follow-up are often developed from surveillance networks which feed databases. More specifically, in France the national occupational disease surveillance and prevention network (Réseau National de Vigilance et de Prévention des Pathologies Professionnelles, RNV3P) is a nation-wide network of experts concerned with diagnosing and monitoring occupational health problems. Any association of a disease with a set of occupational exposures that are potentially causative is entered in the network database. One of the objectives of this network is to develop statistical and modeling methods for study of the structure and the change over time of disease-exposure relationships and to detect emerging disease-exposure associations. With this aim in mind, systematic data-mining methods based on those used in pharmacovigilance have been applied in order to generate pre-alerts concerning statistically significant excess numbers of disease-exposure pairs. In parallel, we have begun to develop an approach that differs from but is complementary to pharmacovigilance methods. This approach takes composite occupational exposures into account in their entirety when describing disease-exposure associations and generates hypotheses as to their cause (Faisandier *et al.* 2007). This latter network approach, based on the exposome or relationship network of occupational health problems, is similar to a number of networks that can be found in the literature, for example interactomes in molecular biology (Chautard *et al.* 2008)**,** or social networks (Barabási *et al.* 2007).

Based on the observations of a network of experts, the RNV3P data do not allow conventional epidemiological analyses, such as estimation of the incidence or prevalence of certain occupational diseases in population groups. The aim of this paper is to present the basis of the exposome approach and to illustrate its potential for the study and programmed surveillance of composite disease-occupational exposure associations (Faisandier *et al.* 2007). In this perspective, we construct a theoretical framework and a formalism to define the space and the principal variables of the system within which the RNV3P data are studied and analysed. We then define and detail the phases of construction of the exposome. As a first illustration of use of this approach, we show how the exposome can define and identify the exposure groups made up of cases that have been subjected to partially or totally identical exposures but which present different diseases. As a second illustrative example, we apply the exposome approach to malignant tumours, focusing on the subgroup of non-Hodgkin lymphomas.

**2. The RNV3P, a database of occupational health problems**

The Réseau National de Vigilance et de Prévention des Pathologies Professionnelles (RNV3P) brings together the 30 national occupational health consultation centres (Centres de Consultation de Pathologies Professionnelles, CCPP) throughout France, where experts in occupational health see patients referred for diagnosis of a potentially work-related disease and which in most cases is of toxicological origin. This network thus forms a database on occupational health problems (Bonneterre *et al.*, in preparation).

An occupational health problem (OHP) is defined as the association of a disease and a composite occupational exposure, of which one or several components are potential causes of the disease (**Fig. 1**). In a structured and standardised manner, the medical experts code a principal disease (ICD-10 classification) and a composite occupational exposure given as $\boldsymbol{e} = (\boldsymbol{n},o,s)^T$, characterised by a set of exposures $\boldsymbol{n}$, an occupation $o$ and an activity sector $s$. The exposure vector $\boldsymbol{n} = (n_1,n_2,n_3,n_4,n_5)^T$ includes 1 to 5 exposures that are suspected or confirmed to cause the symptoms and that are present in the patient's work environment. These exposures are coded using a hierarchical code owned by the national social security organism (Caisse Nationale d'Assurance Maladie, CNAM). The patient's professional activity is described by occupation $o$, coded according to the Classification Internationale Type des Professions (CITP-88) the French equivalent of the International Standard Classification of Occupations (ISCO), and by activity sector $s$, coded according to the French occupational classification (Nomenclature des Activités Professionnelles, NAP-03).

Since 2001, the occupational health consultation centres (CCPP) record an average of 6000 OHP each year, involving occupational diseases. This database is characterized by a four-dimensional space, defined by the system variables (disease, exposures, occupation, sector) according to the codes available in the classifications (**Fig. 2**). The total number of possible separate codes is 1716 for diseases, 6722 for exposures, 390 for occupations and 61 for activity sectors. While occupations and activity sectors are well represented, the disease and exposure codes used cover only a small part of the available space, or 38.96% and 17.70% since 2002 and 6.72% and 3.45%, respectively, each year.

For example (**Table 1**), we find in the database 6501 cases of pleural plaque (code J92) associated with 109 separate exposures for 243 separate occupations and 56 separate activity sectors. Malignant neoplasms of the bronchus and lung (code C34) are associated with 305 separate exposures, 289 occupations and 58 activity sectors, and asthma (code J45) is associated with 524 exposures for 249 occupations and 58 activity sectors. This illustrates the variability of the occupational exposures found in the network according to the diseases observed.

## 3. Development of the exposome

Description of the spectrum of diseases caused by the exposures lies at the heart of the complexity of the relationships between diseases and occupational exposures. Patients diagnosed with a same disease can in fact have been subject to diverse exposures and have carried out a variety of occupational activities. Inversely, patients diagnosed with different diseases may be associated with identical professional exposures. In this context of disease surveillance, the disease diagnosed by the medical experts is the element which triggers the creation of a new recording of an OHP in the RNV3P database. Approaching OHP through the disease is thus coherent with the medical culture of the members of the network, who determine the diagnosis by means of a technical plateau and specialised advice.

The RNV3P database is constructed on a conventional tabular model, where patients are described online by variables-attributes in columns (observation year, disease, exposures, occupation, activity sector). Conventional analysis of this information would consist of examining the distributions and correlations of these variables among individual subjects. However, because of the large number of modalities per variable-attribute and the variable length of the corteges of associated exposures, a conventional statistical approach cannot be used. To study these multidimensional composite disease-exposure relationships, we have chosen a different approach which addresses not only correlations between variables (in columns), but rather emphasizes comparison between cases or patients in order to seek similarities in associated related exposures that may potentially have led to the same diseases. Based on these reflections, we developed the concept of the exposome defined as a network of OHP that have in common one or several elements of occupational exposure. Processing these data in the form of a network of relationships offers wide possibilities of analysis and offers a new way of approaching the data.

In order to do so, each OHP in the RNV3P database is represented by a node (vertex) $v = (p,e)^T$, which is a unique combination of a disease $p$ and its composite associated occupational exposure $e$. Each node is characterised by a weight $w$ equal to the total number of copies of identical OHP in the database (**Table 2**). This representation of the data replaces the initial database by another smaller one, consisting only of separate OHP, i.e. the nodes. Construction of the exposome requires preliminary definition of a rule of connection (a two-way link) between the nodes (separate OHP) reflecting the fact that they have in common one or several components of respective exposures.

We define two nodes $v_i$ and $v_j$ as being connected if $C_{ij,\eta} > 0$ with $C_{ij,\eta} = \sum_{k,l=1}^{5} I(n_{ik}; n_{jl}) \times \theta(w_i - \eta) \times \theta(w_j - \eta)$, where the function $I$ compares the elements of the corteges of exposures $n_i$ and $n_j$ of exposures $e_i$ and $e_j$ in such a way that $I(n_i; n_j) = 1$ if one of the elements of $n_i$ is equal to one of the elements of $n_j$ and $I(n_i; n_j) = 0$ if it is not, and the product $\theta(w_i - \eta) \times \theta(w_j - \eta)$ ensures that the numbers of copies of OHP in each node $v_i$ and $v_j$ are simultaneously greater than a threshold η where $\theta(z) = 1$ if $z \geq 0$ and $\theta(z) = 0$ for $z < 0$.

On this basis, we can then define the $D_\eta$ – exposome as the OHP network that has in common at least D elements of exposure and where each OHP appears at least η times in the database. Such an exposome is represented by the graph $G = \{W,V,L,D, \eta\}$ made up of a set of W OHP described by V nodes connected between themselves by L links and characterised by the adjacency matrix defined by $A_{ij} = 1$ if $C_{ij,\eta} \geq D$ and by $A_{ij} = 0$ if it is not, with i,j∈ {1,…V}. The minimum weight $\eta$ per node allows consideration only of nodes containing at least η OHP (η = 1, by default) and the parameter D indicates the minimum number of common elements of exposure (D = 1, by default).

One of the first ways in which we can use the exposome thus built is to identify the exposure groups defined as a set of OHP (nodes) sharing an identical element of exposure. This is termed a single exposure group, and takes the name of the element of exposure. For this element of exposure, the OHPs of the exposure group represent a potential spectrum of effect.

Our definition of the exposure group is similar to that of a clique in the language of graph theory. Strictly speaking, a clique is a set of *n* nodes connected by exactly *n–1* links with the other nodes contained in the clique. This definition of the clique does not place any restriction on the nature of the links. Where exposomes are concerned, cliques may form « hybrid » exposure groups defined as a set of *n* nodes connected by exactly *n-1* links which may be of different kinds (several elements of the exposure are shared).

In the same way, study of networks (social networks in particular) makes use of a certain number of statistical measurements or metrics to describe groups of individuals that are formed by characterising the properties relative to the structure of the system (Barabási 2007; Park *et al.* 2007). The simple metrics most frequently used are density d ($d = 2L/V(V-1)$, the probability that two nodes taken at random are connected, distribution P(k) of connectivity degrees k (number of links or connections per node), and clustering coefficients c (probability that two nodes connected to the same node are also connected between themselves) (Watts *et al.* 1998). We shall illustrate the interest of these metrics and exposure groups by taking the exposome of non-Hodgkin lymphomas (NHL) as an example. The graphs (**Fig. 3** and **Fig. 4**) were created using Ucinet and NetDraw software (Borgatti *et al.* 2002).

**4. Illustration**

*4.1. Exposome of malignant tumours reported in the RNV3P database*

In order to illustrate what an exposome may look like, we considered the subset of malignant tumours which concerns 3990 OHP of potentially occupational origin in the RNV3P database (**Fig. 3**) for the period 2002-2007. We identified 195 separate associations (nodes) between a type of malignant tumour (node colour) and a cortege of 1 to 5 associated exposures. Most types of malignant tumours observed affect the respiratory system, notably malignant tumours of the bronchus and lung (65% of malignant tumours, or 2606/3990) and mesotheliomas (13%, or 532/3990). Among the diseases that accounted for more than 5% of observations, we found 210 malignant tumours of the bladder (classified as malignant urogenital tumours).

The $1_1$ - exposome of malignant tumours (obtained by taking the nodes containing at least $\eta = 1$ OHP and linked by at least $D = 1$ common exposure) is not a very dense network (d = 0.20) with 211 separate exposures shared between the nodes. The central exposure group is a highly connected group, composed of nodes linked between themselves by exposure to asbestos and consisting mainly of respiratory diseases. This exposure group contains nodes connected by common exposures other than asbestos. As for the $2_{10}$ – exposome (obtained by taking the nodes containing at least $\eta = 10$ OHP and linked by at least $D = 2$ common exposures), it represents nearly three-quarters of the patients (2648/3990) described by 25 nodes. This exposome presents another dimension included in the structure of the $1_1$ – exposome. Among the nodes of these exposomes, we find for example lymphomas.

Lymphomas belong to unconnected exposure groups which share exposures with other types of malignant tumours. Among the exposures associated with lymphomas, we find asbestos, polycyclic aromatic hydrocarbons, organic solvents and thinners, benzene, agricultural products and ionising radiations. By closer analysis of the detailed structure of the exposome built from the subgroup of lymphomas, we can characterise the exposures found in the occupational environment of the patients and study the complexity of their interactions.

*4.2. Detailed structure of the exposome of malignant tumours: a closer view of non-Hodgkin lymphomas*

Non-Hodgkin lymphomas (NHL) are malignant tumours of the lymphatic system. A review of the literature on environmental and professional factors (Alexander *et al.* 2007) does not make it possible to identify specific exposures systematically associated with NHL. The RNV3P database on these lymphomas reflects this state of knowledge. The sample extracted from the RNV3P brings together 77 OHP. The male/female ratio is 4.5 (63/14) and mean age is 52.3 years (52.0 for men and 53.7 for women). Median age for both sexes is 54 years. This sample presents 72 separate exposure codes associated with NHL (54 codes after aggregation), 55 occupation codes and 28 different activity sector codes. Nearly half the OHP reported (38/77) are associated with several exposures.

The $1_1$ - exposome of NHL (**Fig. 4**) consists of the 77 OHP (W = 77) distributed in 51 nodes ($V = 51$), of which 46 nodes share at least one exposure. It represents the spectrum of exposures associated with patients with NHL. Each node represents *w* patients associated with a set of exposures present in the occupational environment and potentially causative. For example, one patient had been exposed to both carboxylic acid and peracids as well as to antineoplastic drugs. Isolated nodes are also observed, in particular the case of a patient exposed to wood dust who did not share any other exposure with the other patients.

*4.3. Structure of the NHL exposome*

Twenty-four groups of single exposures were identified in the $1_1$ - exposome of NHL. For example, the group of organic solvents and thinners is made up of 15 nodes (unique combinations between an NHL and a cortege of associated exposures). At this stage, we can then reduce the $1_1$ - exposome of NHL by an equivalent exposome of the exposure groups. Such an exposome shows the connected structure of the spectrum of exposures that are potentially causative of NHL. Another way of representing this aspect of the question is to construct a dendogram.

**Fig. 5** presents a dendrogram of the interlinked structure of the 24 exposure groups associated with NHL. This dendrogram is constructed from the adjacency matrix. For example, the exposure groups benzene and halogen derivatives of saturated aliphatic hydrocarbons are linked by two nodes associated with these two exposures. These two exposure groups are themselves connected to the group of exposure to black products, which is itself connected to the ionising radiation exposure group, and so on. Twenty-two hybrid exposure groups (cliques) can also be identified. For example, 3 nodes form the hybrid exposure group « petrol x lubricating oils and greases x organic solvents and thinners ». Each of these carries exactly 2 connections within the clique but they do not share the same exposures: 2 share petrol exposure, 2 exposure to lubricating oils and greases and 2 exposure to organic solvents and thinners.

As a node is associated with a maximum of 5 exposures, it can therefore share up to 4 exposures with other nodes. In the exposome, this is reflected by overlapping or interlinking of exposure groups. For example, two patients were exposed to black products, while one was also exposed to organic solvents and thinners and another to benzene. These two nodes thus make up the group of those exposed to black products, and are also part of another exposure group (organic solvents and thinners, and benzene, respectively). These observations lead to overlapping of exposure groups, demonstrating the complex relationships between exposure factors.

More rational than the visual approach, we can use the connectivity degree and the clustering coefficient to study the complex architecture of the exposome. **Fig. 6** illustrates, for each connected node, the distribution of connectivity degrees k, classified in decreasing order. For each value of k, we observe values of the clustering coefficient of c which vary between 0 (the probability that all adjacent nodes are connected is nil) and 1 (all adjacent nodes are interconnected). The nodes with the most connections are associated with the exposure organic solvents and thinners where k ranges from 14 links (the node « NHL x organic solvents and thinners ») to 23 links (the node « NHL x organic solvents and thinners x ionising radiations x benzene »).

The values of the clustering coefficient of c determine the properties of the nodes within the exposome. The connected nodes which belong to only one exposure group are characterised by c = 1, unlike those belonging to at least two groups where $0 < c < 1$, and singletons where c = 0.

A relationship is observed between connectivity degree k and clustering coefficient c within the exposure group of organic solvents and thinners, although this does not allow us to conclude on the systematic existence of a k - c relationship, nor to define the true incidence of this relationship on the topology of the network. However, we can identify nodes characterised by low values of k and c as « bridging nodes ». These nodes form a junction between 2 to 4 exposure groups through the shared exposures ($D_{max} = 4$). For example, the node « NHL x toluene x virus x aliphatic aldehydes » links the 3 exposure groups.

*4.4. Variation of the exposome over time*

Between 2002 and 2006, 62 observations were reported (in white in **Fig. 4**). During the year 2007, 15 new OHP were recorded by the medical experts and 10 of these corresponded to new associations which had never previously been described (black nodes), while 5 which had already been observed during the period 2002-2006 (increased numbers within a node). Projection of the 2007 data on the exposome of the cumulative data reveals new events (black nodes and increased numbers of observations) which modify the structure of the exposure groups (see electronic supplementary material).

*4.5. Projection of the $1_1$ - exposome of NHL on occupational activity*

The exposome identifies groups of patients subjected to the same exposures, but do all these groups have the same occupational activity ? From a probabilistic viewpoint, the most plausible hypothesis would be to observe exposure groups composed of patients who had had similar or related occupational activities. In the graphic representation of the exposome, the location of the observations associated with identical occupations or activity sectors reveals the variability of exposures.

As an example, the occupations « mechanical engineering technicians » (5 cases: 1 in 2002, 1 in 2004, 1 in 2005, 1 in 2006 and 1 in 2007, black squares) and « motor vehicle mechanics and fitters » (4 cases: 1 in 2006 and 3 in 2007, black circles) appear on the exposome. These occupations belong to different activity sectors but share similar exposures, to benzene in particular, but also to other solvents.

The association of these exposures with the diagnosis of NHL, common to both occupations, can then be investigated. Although methods of detecting emerging disease-exposure pairs do not specifically pinpoint these occupations by testing the exposures separately, the exposome does generate a new hypothesis with regard to the relationship between NHL and solvents according to type of occupation.

**5. Concluding Remarks**

We have developed the theoretical framework and the concept of the exposome, defined as a network of OHP linked by similar exposures, which makes it possible to study and to analyse the structure and typology of the disease-occupational exposure associations reported in the RNV3P database. As illustrated by the example of malignant tumours and NHL, the exposome enables us to identify exposure groups, or the set of patients exposed to at least one identical exposure, and to study the complexity of the interactions between these groups in the overall structure of the network.

Whereas analytic epidemiology traditionally aims to test hypotheses that have been advanced beforehand, the exposome approach is able to supply, by mapping the spectrum of exposures, a typology of expert-reported knowledge that cannot be observed by conventional statistical methods. Organisation of this knowledge makes it possible to formulate hypotheses based on resemblances between OHP in terms of occupational exposures associated with a disease.

In the literature, we find that the concept of the exposome has been used to examine several types of problem. For example, in a study of the spread of obesity, Christakis *et al.* used the concept of relationship networks by connecting individuals who shared a similar social network in order to follow change in their weight over time (Christakis *et al.* 2007). Starting from the hypothesis that an individual's weight gain may be influenced by his or her social contacts, this approach revealed that the social network could be a factor in the spread of the obesity epidemic. Pursuing this idea, Barabási conceptualised this network using the term of « diseasome » (Barabási 2007). The social network of obese persons, for example, may be linked with a wider network of exposures, and all these factors can be studied collectively in order to take into account the combination of components which could have a causative effect on health. In another field, the term of exposome has already been used in molecular genetics to construct a network of environmental exposures shared by individuals (Wild, 2005). The objective of this network is to better understand and describe the role of each of the factors in health and so to generate causative hypotheses. Nevertheless, this approach requires a certain amount of information on the tracability of the exposures, but it also lays the basis for coherent reflection on the development of relational networks for surveillance of composite exposures of various origins. In the same spirit, other authors have begun to reason in terms of « human disease networks », which are networks of human diseases or disorders that share similar genetic mutations, while the patients do not necessarily develop the same disease. The aim of this approach is to generate hypotheses as to the roles of genetic mutations in the development of diseases (Goh *et al.* 2007).

All these examples show that it is the search for similarities between situations that are apparently different that guides all subsequent analyses. In our context, it is the construction of the exposome that serves as the starting point for analysis, as we have shown in our illustration of identification of exposure groups. We should remember that unlike epidemiological methods which lead to calculations of prevalence or risk estimations, the exposome approach we develop here is applied to analysis of composite disease-occupational exposure associations with the aim of health surveillance. Similarly, compared with pharmacovigilance methods that can detect disease-exposure pairs which produce an emerging signal, the exposome offers a complementary approach to be developed as a tool for the surveillance over time of diseases associated with composite exposures. With nearly 60,000 OHP presenting as occupational diseases reported since 2001, and for which composite occupational exposures have been identified, we may hope that potentially emerging disease-exposure associations may soon be detected.

Finally, the exposome approach opens new horizons and we have hardly begun to explore the vast possibilities of information that it offers. The work presented in this paper can in fact be developed in several directions, for example by systematic study of distributions of connectivity degrees and clustering coefficients, and by study of the dynamics of exposomes. In addition, without being restricted only to exposures as we have presented in this paper, exposomes can be analysed at several levels: at the level of a single disease or of all reported exposures, taking into account the cortege of exposures, occupations and activity sectors. Each scale of analysis has its corresponding exposome, providing information on the exposure. Analysis at the level of several different diseases would result in an interlocking, hierarchical structure of exposomes. All these aspects of the question of relationships and disease-exposure associations will be examined in a forthcoming work.

| **GLOSSARY** |
|---|
| **OHP:** Occupational health problem resulting from the association of a disease and of a potentially composite occupational exposure, of which one or several components are potentially causative (exposures, occupation and/or activity sector). |
| **Node:** Combination or single association between a disease and its associated composite occupational disease exposure. |
| **Exposure group:** Set of OHPs which share an identical element of exposure. |
| **Clique:** Set of *n* nodes connected by exactly *n-1* links with the other nodes. |
| **Density:** Probability that two nodes taken at random are connected (ratio of the number of connections of the network to the maximum number of possible links) |
| **Connectivity degree:** Number of links or connections per node. |
| **Clustering coefficient:** Probability that two nodes connected to a given node are also connected between themselves. |
| $D_\eta$ **– exposome:** Network of OHPs which have at least D elements of exposure in common, and in which each OHP appears at least $\eta$ times in the database. |

ACKNOWLEDGEMENTS

The authors thank the Agency for Environmental and Occupational Health Safety (AFSSET) for supporting this work, the national health insurance organisation (Caisse Nationale d'Assurance Maladie, CNAM) for funding of the consultations and participation in funding of the RNV3P, the engineers of the regional branches of the national health insurance organisation (Caisses Régionales d'Assurance Maladie, CRAM), , Sylvette Liaudy for her help on bibliography, Lynda Larabi for her contribution to the processing and quality control of data, the staff of the occupational health consultation centres who supply the RNV3P with data, the sentinel physicians of the occupational health services for recording and transmitting incident occupational health reports, and Nina Crowte for assistance in the translation from French of the manuscript.

REFERENCES

Alexander, D.D., Mink, P.J., Adami, A.O., Chang, E.T., Cole, P., Mandel, J.S. & Trichopoulos, D. 2007 The non-Hodgkin lymphomas: a review of the epidemiologic literature. *Int J Cancer* **120** Suppl 12 1-39. (DOI 10.1002/ijc.22719)

Barabási, A. L. 2007 Network medicine – from obesity to the "diseasome". *N Engl J Med* **357(4),** 404-407.

Barabási, A.L. 2007 The architecture of complexity. *IEEE Control Systems Magazine* **27(4)**, 33-42.

Blair, A. Rothman, N. & Zahm, S.H. 1999 Occupational cancer epidemiology in the coming decades. *Scand J Work Environ Health*, **25(6)**, 491-497.

Bonneterre, V., Bicout, D.J., Larabi, L., Bernardet, C., Maitre, A., Tubert-Bitter, P. & de Gaudemaris, R. 2008 Detection of emerging diseases in occupational health: usefulness and limitations of the application of pharmacosurveillance methods to the database of the French National Occupational Disease Surveillance and Prevention network (RNV3P). *Occup Environ Med*, **65**, 32-37.

Bonneterre, V., Faisandier, L., Bicout, D., Bernardet, C., Ameille, J., De Clavière, C., Aptel, M., Lasfargues, G. & de Gaudemaris, R. 2008 Programmed health surveillance and detection of emerging diseases in occupational health: contribution of the French national occupational disease surveillance and prevention network (RNV3P). *Occup Environ Med* (currently submitted).

Borgatti, S.P., Everett, M.G. & Freeman, L.C. 2002 *Ucinet for Windows: Software for Social Network Analysis*. Harvard, MA: Analytic Technologies.

Chautard, E., Thierry-Mieg, N. & Ricard-Blum S. 2008 Interaction networks: From protein functions to drug discovery. A review. *Pathol Biol*, **57**, 324-333.

Christakis, N.A. & Fowler J.H. 2007 The spread of obesity in a large social network over 32 years. *N Engl J Med*, **357**, 370-379.

Faisandier, L., Bonneterre, V., de Gaudemaris, R. & Bicout, D.J. 2007 Elaboration d'une méthode statistique pour la détection d'événements émergents: application au Réseau National de Vigilance et de Prévention des Pathologies Professionnelles (RNV3P). *Epidémiol et Santé Anim*, **51**, 111-118.

Goh, K.I., Cusick, M.E., Valle, D., Childs, B., Vidal, M. & Barabási, A.L. 2007 The human disease network. *Proc Natl Acad Sci U S A,* **104**, 8685-8690. (DOI 10.1073/pnas.0701361104.)

Park, J. & Barabási, A.L. 2007 Distribution of node characteristics in complex networks. *Proc Natl Acad Sci U S A,* **104**: 17916-17920. (DOI 10.1073/pnas.0705081104.)

Watts, D.J. & Strogatz, S.H. 1998 Collective dynamics of 'small-world' networks *Nature*, **393**, 440-442.

Wild, C.P. 2005 Complementing the genome with an "exposome": the outstanding challenge of environmental exposure measurement in molecular epidemiology. *Cancer Epidemiol Biomarkers Prev*, **14**, 1847-1850 (DOI 10.1158/1055-9965.EPI-05-0456.).

TABLES

Table 1. Characteristics of the diseases most often represented in the RNV3P database (n = 57,368 OHP, presenting as work-related diseases, reported between 2001 and 2007; 813 separate diseases associated with 1304 separate exposures, 385 separate occupations and 62 separate activity sectors).

| Disease (ICD-10 codes) | Number of observations in RNV3P | Number of separate exposures | Number of separate occupations | Number of separate activity sectors |
|---|---|---|---|---|
| Pleural plaque (code J92) | 6501 (11.3%) | 109 (8.3%) | 243 (63.1%) | 56 (90.3%) |
| Malignant neoplasms of bronchus and lung (code C34) | 3821 (6.6%) | 305 (23.4%) | 289 (75.1%) | 58 (93.5%) |
| Asthma (code J45) | 3556 (6.2%) | 524 (40.2%) | 249 (64.7%) | 58 (93.5%) |

Table 2. Descriptive parameters of the exposome.

| | Parameters relating to an OHP |
|---|---|
| $p$ | Principal disease |
| $\boldsymbol{e}$ | Composite occupational exposure |
| $\boldsymbol{n}$ | Cortege of associated exposures |
| $o$ | Job involved |
| $s$ | Activity sector involved |
| | Parameters relating to a node $\boldsymbol{v}_i$ |
| $w_i$ | Total number of copies of node i in the database |
| | Parameters relating to exposome $G$ |
| $W$ | Size of the sample considered |
| $V$ | Total number of separate nodes |
| $L$ | Total number of links |
| $D$ | Minimum number of shared exposures |
| $\eta$ | Minimum weight of the nodes |

Figure 1. Structure of an occupational health problem

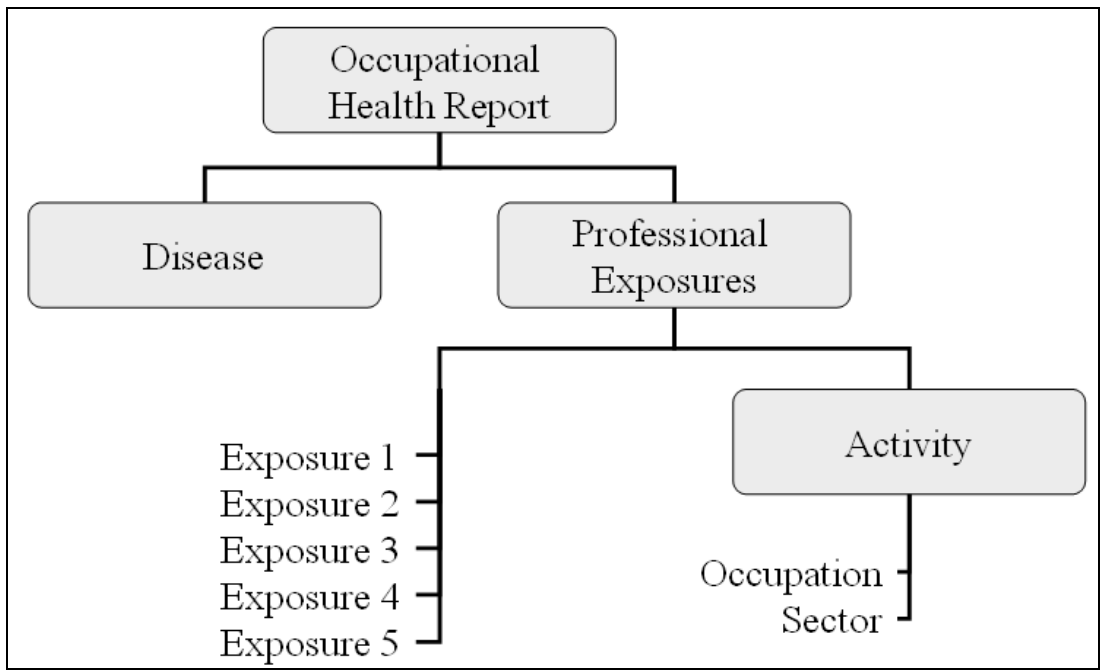

Figure 2. Polar graph showing the proportion of space taken up by the codes used among the codes available. Data compiled from the OHP observed during the period 2002-2007.

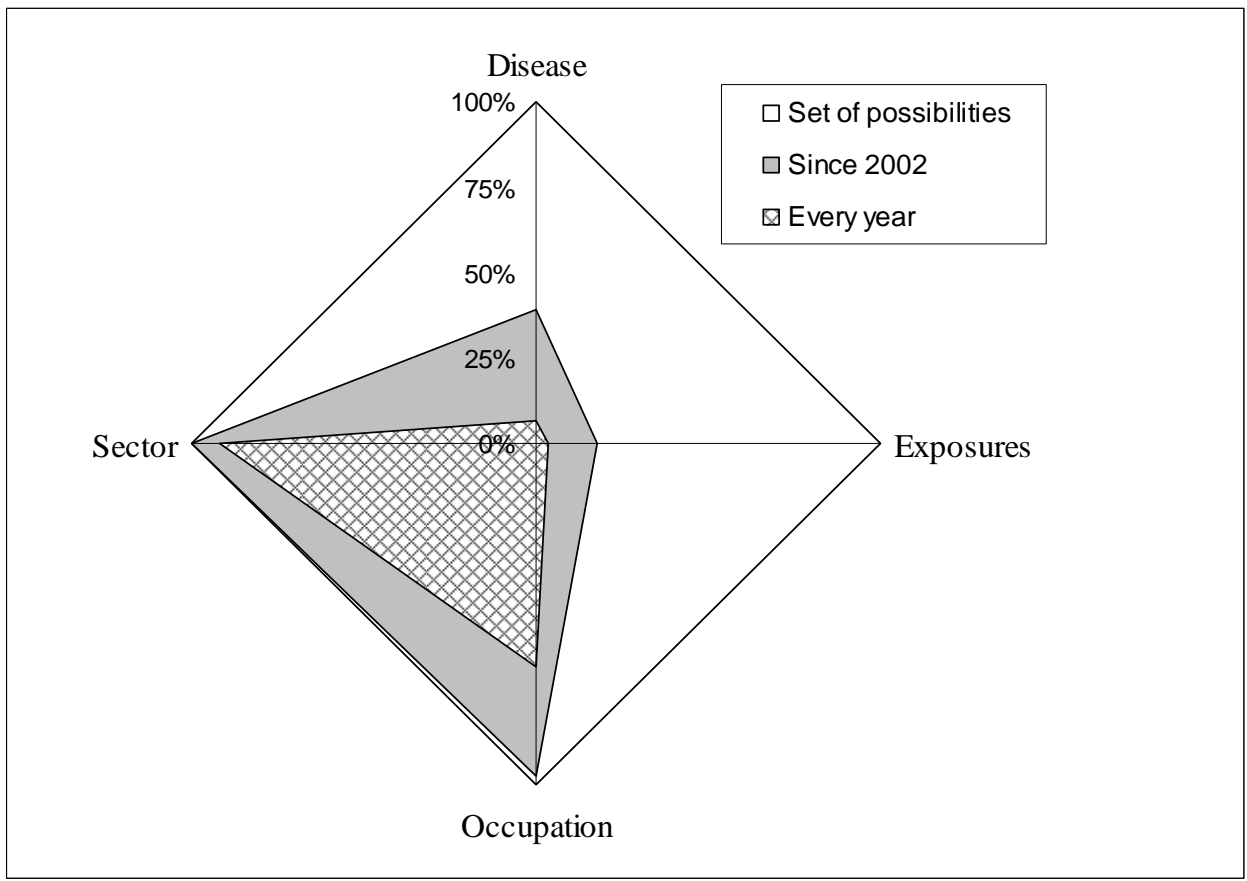

Figure 3. Exposomes of malignant tumours recorded in the RNV3P database during the period 2002-2007.
  (a) The $1_1$ – exposome with $G_{TM}$ = {W=3990, V=195, L=3715, D=1, $\eta$=1}. Each node represents a type of malignant tumour, colour coded, associated with a cortege of 1 to 5 exposures. The size of the nodes is proportional to the number of identical OHP reported. The central group of exposure represents, in part, nodes associated with asbestos and other possible co-exposures. Four nodes are composed of more than 50 OHP: a) 58 malignant tumours of the bronchus and lung are associated with exposures to asbestos x polycyclic aromatic hydrocarbons, b) 45 malignant tumours of the bronchus and lung and 5 mesotheliomas are associated with exposures to asbestos x welding fumes, c) 49 bronchopulmonary cancers and 2 mesotheliomas are associated with asbestos x crystallised silica, and d) 65 malignant tumours of the facial sinuses and 3 malignant tumours of the bronchus and lung are associated with wood dust.
  (b) The $2_{10}$ – exposome with $G_{TM}$ = {W=2648, V=25, L=2, D=2, $\eta$=10} reveals only the 25 nodes which share at least 2 exposures and are composed of 10 or more OHP. Three nodes are linked by exposures to asbestos x PAH (polycyclic aromatic hydrocarbons) and to silica sand x asbestos.

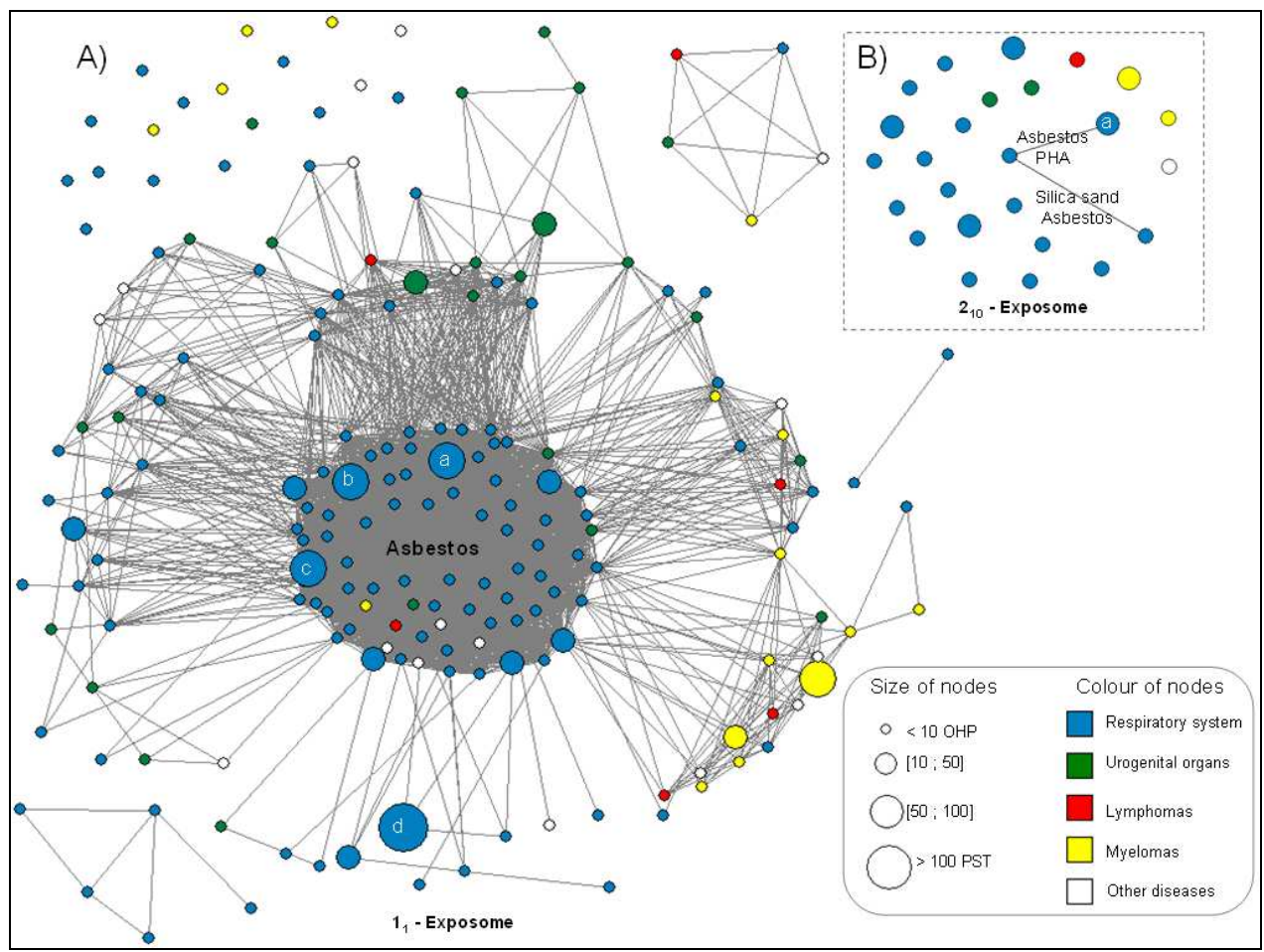

Figure 4. The $1_1$ – exposome of non-Hodgkin lymphomas observed during the period 2002-2007, where $G_{NHL} = \{W=77, V=51, L=210, D=1, \eta=1\}$. A node represents an NHL associated with a cortege of 1 to 5 exposures and contains at least 1 disease-exposure association. The nodes are connected if they share at least one exposure. The size of the nodes is proportional to the number of OHP. The numbers of OHP for each exposure group are given in brackets (for the whole period 2002-2006, with the increase in numbers for 2007). Co-exposures that are not shared do not appear in the chart.

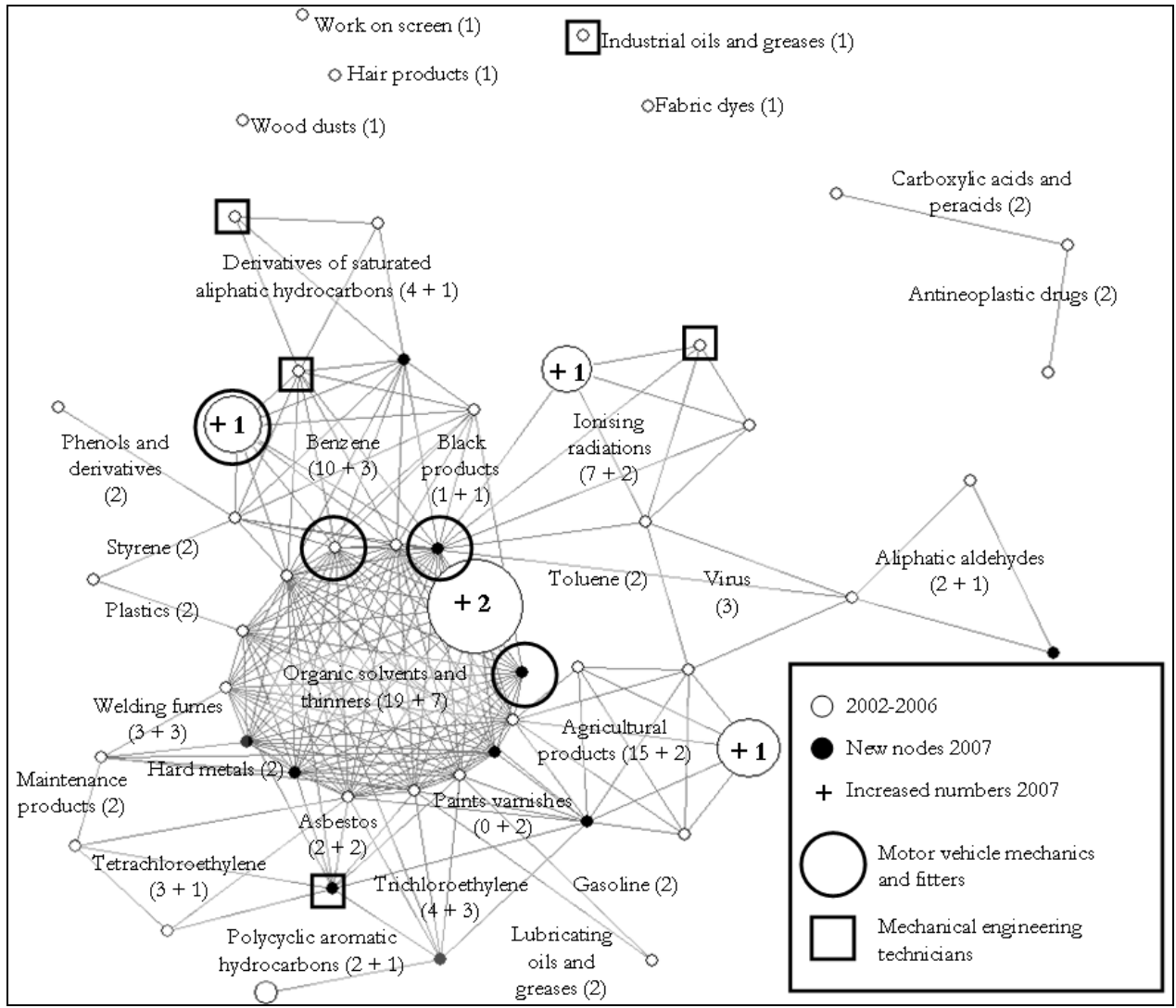

Figure 5. Hierarchical structure of exposure groups, originating from the $1_1$- exposome of NHL, shown as a dendrogram. The arborescence reflects the interlinked organisation of the exposure groups within the exposome and the proximity of these groups can thus be observed (average linkage method, R 2.8.1 software). The numbers of OHP by exposure group for the overall period 2002-2007 are shown in brackets. The leaves of the tree form 24 exposure groups and the root of the tree represents the global exposome, i.e. the range of shared exposures associated with NHL.

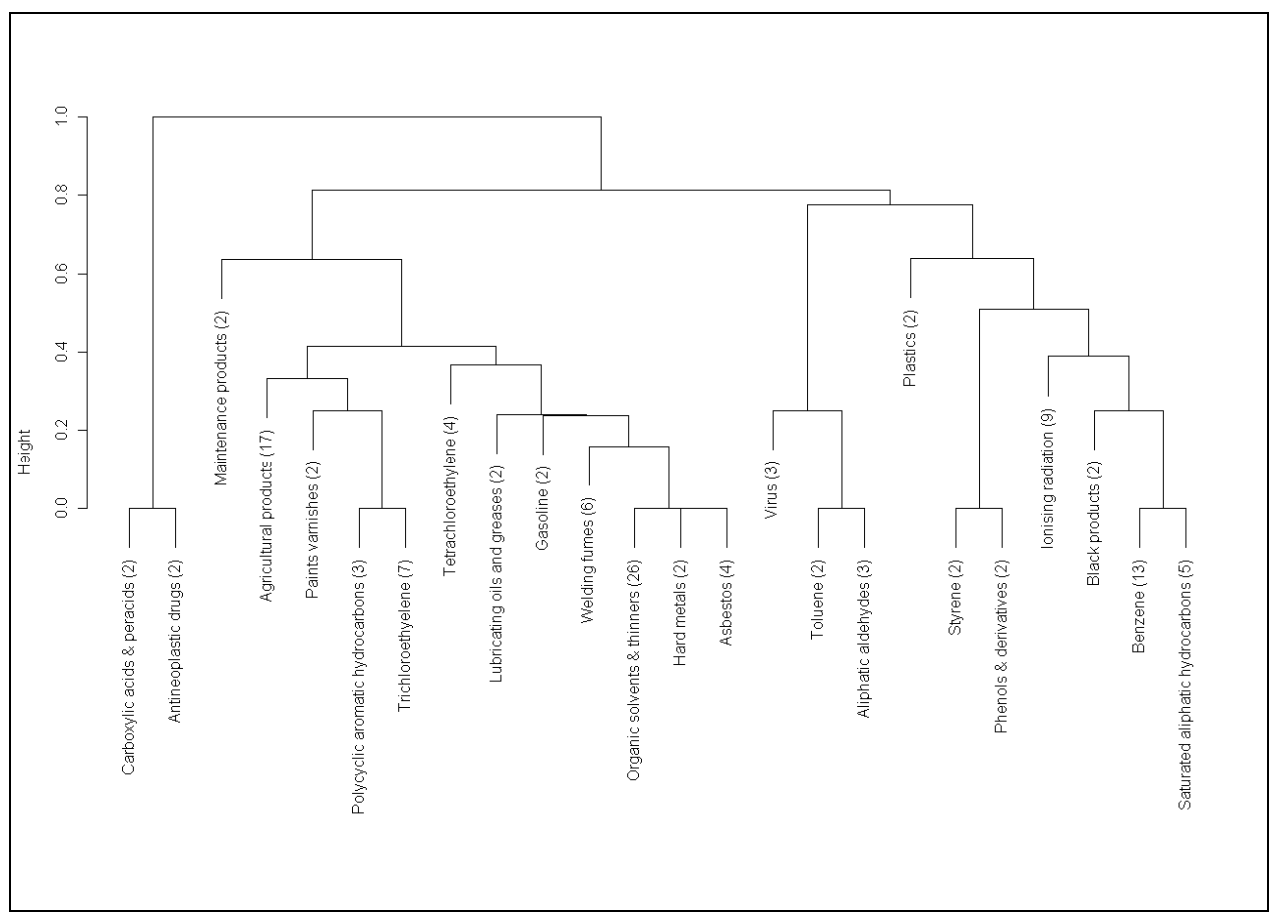

Figure 6. Distribution of connectivity degrees (number of connections per node) and clustering coefficients (probability that the adjacent nodes are also interconnected) for nodes with more than one connection.

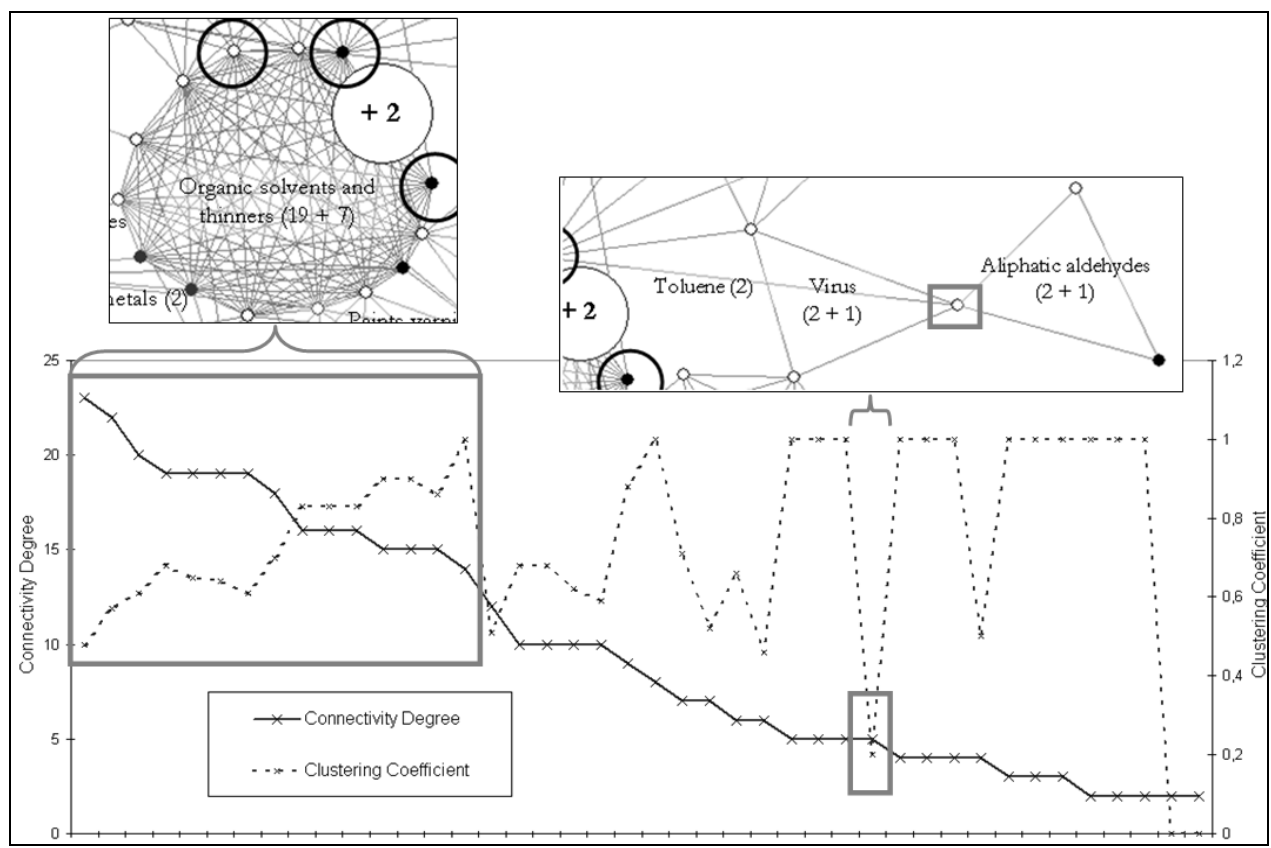

ELECTRONIC SUPPLEMENTARY MATERIAL

Table A. List of exposures (by aggregate codes) associated with NHL, classified by main groups of exposures (1st digit) and then by numbers in each group.

| Associated exposures | Observation year | | | | | | TOTAL |
|---|---|---|---|---|---|---|---|
| | 2002 | 2003 | 2004 | 2005 | 2006 | 2007 | |
| 1 - Inorganic compounds | | | | | | | |
| Bromhydric acid, hydrogen bromide | 1 | | | | | | 1 |
| Fluorhydric acid, hydrogen fluoride | | | | | | 1 | 1 |
| Mercury | | | | | 1 | | 1 |
| 2 - Organic compounds | | | | | | | |
| Benzene | 2 | 1 | 2 | 1 | 4 | 3 | 13 |
| Trichloroethylene | | | 1 | 3 | | 3 | 7 |
| Halogen derivatives of saturated aliphatic hydrocarbons | 1 | | 3 | | | 1 | 5 |
| Tetrachloroethylene | | 1 | | 1 | 1 | 1 | 4 |
| Aliphatic aldehydes | | | | 2 | | 1 | 3 |
| Polycyclic aromatic hydrocarbons | | 1 | | 1 | | 1 | 3 |
| Phenols and phenol derivatives | 1 | | | 1 | | | 2 |
| Toluene | | | | 1 | 1 | | 2 |
| Styrene | 1 | | | | 1 | | 2 |
| Carboxylic acids and peracids | 1 | | 1 | | | | 2 |
| Monocyclic aromatic hydrocarbons | | | | 1 | | | 1 |
| Methanol | 1 | | | | | | 1 |
| Diethylene glycol monobutyl ether | | | | | 1 | | 1 |
| Other aromatic aldehydes | 1 | | | | | | 1 |
| Aliphatic ketones | 1 | | | | | | 1 |
| Other thioacids | | | | | 1 | | 1 |
| Glycol esters | 1 | | | | | | 1 |
| Acrylamide | 1 | | | | | | 1 |
| Cyanates and isocyanates | | | | 1 | | | 1 |
| Aromatic amines and derivatives | | | | | | 1 | 1 |
| Aromatic nitro derivatives | | | | 1 | | | 1 |
| 3 – Industrial substances | | | | | | | |
| Organic solvents and thinners | 6 | 1 | 2 | 5 | 5 | 7 | 26 |
| Agricultural products | | 5 | 3 | 3 | 4 | 2 | 17 |
| Welding fumes | | | | 2 | 1 | 3 | 6 |
| Asbestos | | | 1 | 1 | | 2 | 4 |
| Paints, varnishes, lacquers, putties | | | | | | 2 | 2 |
| Plastics, polymer rubbers | 1 | | | | 1 | | 2 |
| Hard metals, metallic carbons | | | | | | 2 | 2 |
| Black products (tars, bitumen, asphalts) | | | 1 | | | 1 | 2 |
| Petrol | | | 2 | | | | 2 |
| Lubricating oils and greases | | | 1 | 1 | | | 2 |
| Cleaning products | | | | | 2 | | 2 |
| Wood dust | 1 | | | | | | 1 |
| Textiles, vegetal fibres | | | | | | 1 | 1 |
| Synthetic fibres | | | | | | 1 | 1 |
| Diesel gas | | | | | 1 | | 1 |
| Industrial oils and greases | | | | | 1 | | 1 |
| Heat decomposition products | | | | | | 1 | 1 |
| Glues, adhesives | | | | | | 1 | 1 |
| Epoxydic resins | | | | | | 1 | 1 |
| Cloth dyes | | | | | 1 | | 1 |
| Inks | | | | | 1 | | 1 |
| Oils | 1 | | | | | | 1 |
| Leather treatment products | | | 1 | | | | 1 |
| Black product wastes (asphalt, tars, creosote) | | | | | | 1 | 1 |
| Capillary products | 1 | | | | | | 1 |
| **4 – Physical exposures** | | | | | | | |
| Ionising radiations | 2 | 2 | | 1 | 2 | 2 | 9 |
| Work on screen | | 1 | | | | | 1 |
| **7 - Virus** | | | | | | | |
| Virus | | 1 | 1 | 1 | | | 3 |
| **9 – Animal species and substances of animal origin** | | | | | | | |
| Mammals | | | 1 | | | | 1 |
| **C – Medications and drugs** | | | | | | | |
| Antineoplastic drugs | | 1 | 1 | | | | 2 |

Table B. List of occupational codes (maximum 4 digits) associated with NHL classified by major unit group (first digit), and then by number of cases.

| Job titles | Observation year | | | | | | TOTAL |
|---|---|---|---|---|---|---|---|
| | 2002 | 2003 | 2004 | 2005 | 2006 | 2007 | |
| **1 – Legislators, senior officials and managers** | | | | | | | |
| Directors and managers in manufacturing industries | | 1 | | | | | 1 |
| Directors and managers in construction and public works | | | 1 | | | | 1 |
| Directors and managers in wholesale and retail trade | | | | | 1 | | 1 |
| **2 – Professionals, intellectual and scientific occupations** | | | | | | | |
| Biologists, botanists, zoologists and related professionals | 1 | | | 1 | | | 2 |
| Chemists | 1 | | | | | | 1 |
| Civil engineers | | 1 | | | | | 1 |
| **3 – Technicians and associate professionals** | | | | | | | |
| Mechanical engineering technicians | 1 | | 1 | 1 | 1 | 1 | 5 |
| Technicians in physical sciences and chemistry | | | | 1 | 1 | 1 | 3 |
| Technicians in industrial chemistry | 1 | | | | 1 | | 2 |
| Electrical technicians | | | | | | 1 | 1 |
| Physical and engineering science technicians not elsewhere classified | | | | 1 | | | 1 |
| Agronomy and forestry technicians | | | | | | 1 | 1 |
| Medical assistants | | | | | 1 | | 1 |
| Dental assistants | | | | | 1 | | 1 |

|  |  |  |  |  |  |
|---|---|---|---|---|---|
| Pharmaceutical and dispensing assistants |  | 1 |  |  | 1 |
| Nursing staff (intermediate level) | 1 |  |  |  | 1 |
| Special education teaching associate professionals |  |  |  | 1 | 1 |
| **4 – Clerks** | | | | | |
| Stock clerks |  | 1 |  |  | 1 |
| Other office workers | 1 |  |  |  | 1 |
| **5 - Service workers and shop and market sales workers** | | | | | |
| Hairdressers, barbers, beauticians and related workers | 1 |  |  |  | 1 |
| Shop salespersons and demonstrators |  |  |  | 1 | 1 |
| **6 - Skilled agricultural and fishery workers** | | | | | |
| Mixed crop growers |  | 1 | 1 |  | 1 | 3 |
| Tree and shrub crop growers |  |  | 1 |  | 1 |  | 2 |
| Field crop and vegetable growers |  |  | 1 |  |  | 1 |
| Gardeners, horticultural and nursery growers |  |  | 1 |  |  | 1 |
| Market-oriented crop and animal producers |  |  | 1 |  |  | 1 |
| Subsistence agricultural and fishery workers |  |  | 1 |  |  | 1 |
| **7 - Craft and related trade workers** | | | | | |
| Motor vehicle mechanics and fitters |  |  |  | 1 | 3 | 4 |
| Varnishers and related painters |  |  |  | 1 | 1 | 2 |
| Welders and flame-cutters | 1 |  |  | 1 |  | 2 |
| Sheet metal workers-boilermakers |  | 1 |  | 1 |  | 2 |
| Machine-tool setters and operators |  |  | 1 |  | 1 | 2 |
| Compositors, typesetters and related workers |  |  |  |  | 2 | 2 |
| Carpenters and joiners, construction |  |  |  |  | 1 | 1 |
| Floor layers and tile setters |  |  | 1 |  |  | 1 |
| Plumbers and pipefitters | 1 |  |  |  |  | 1 |
| Building and related electricians | 1 |  |  |  |  | 1 |
| House painters and paper hangers | 1 |  |  |  |  | 1 |
| Structural-metal preparers and erectors |  |  |  |  | 1 | 1 |
| Toolmakers and related occupations |  |  |  |  | 1 | 1 |
| Printing engravers, photo-engravers and etchers |  |  | 1 |  |  | 1 |
| Furriers and related workers |  |  |  | 1 |  | 1 |
| Shoe-makers and related workers |  | 1 |  |  |  | 1 |
| **8 - Plant and machine operators and assemblers** | | | | | |
| Chemical still and reactor operators |  |  |  | 1 |  | 1 | 2 |
| Chemical-processing-plant operators not elsewhere classified | 1 |  |  |  |  | 1 |
| Power-production plant operators |  |  | 1 |  |  | 1 |
| Mining-plant operators |  | 1 |  |  |  | 1 |
| Pharmaceutical- and toiletry-products machine operators | 1 |  |  |  |  | 1 |
| Metal finishing-, plating- and coating-machine operators |  | 1 |  |  |  | 1 |
| Textile-, fur- and leather-products machine operators |  | 1 |  |  |  | 1 |
| Electrical-equipment assemblers |  |  |  |  | 1 | 1 |

| Occupation | 2002 | 2003 | 2004 | 2005 | 2006 | 2007 | TOTAL |
|---|---|---|---|---|---|---|---|
| Electronic-equipment assemblers | 1 | | | | | | 1 |
| Metal-, rubber- and plastic-products assemblers | | | | | 1 | | 1 |
| 9 – Elementary occupations | | | | | | | |
| Helpers and cleaners in offices, hotels and other establishments | 1 | 1 | | | | | 2 |
| Hand-launderers and pressers | | | 1 | | 1 | | 2 |

Table C. Activity sector codes (2 digits maximum) associated with NHL.

| Activity sectors | Observation year | | | | | | TOTAL |
| | 2002 | 2003 | 2004 | 2005 | 2006 | 2007 | |
|---|---|---|---|---|---|---|---|
| C – Mining and quarrying | | | | | | | |
| Mining (not specified) | 1 | 3 | 2 | 1 | | 1 | 8 |
| Mining of uranium | | | | | | 1 | 1 |
| D – Manufacturing | | | | | | | |
| Textiles | 1 | | | | | | 1 |
| Fur and apparel | | | | 1 | | | 1 |
| Leather and footwear | | | 1 | | | | 1 |
| Wood and products of wood | | | | | | 1 | 1 |
| Paper and paperboard | | | | | | 1 | 1 |
| Publishing, printing, reproduction | | | | 1 | 2 | | 3 |
| Chemicals | 3 | | | 1 | 1 | 1 | 6 |
| Rubber and plastic products | | | 1 | | 1 | | 2 |
| Manufacture of basic metals | | 2 | | 1 | 1 | 2 | 7 |
| Manufacture of machines and equipment | | | 1 | | | | 1 |
| Manufacture of office machines and computer products | 1 | | | | | | 1 |
| Manufacture of electrical machines and appliances | 1 | | | | | 1 | 2 |
| Manufacture of motor vehicles | | | | | 1 | 1 | 2 |
| Manufacture of other transport equipment | | | | 1 | | 1 | 2 |
| E - Production and distribution of electricity, gas and water | | | | | | | |
| Production and distribution of electricity, gas and heat | | | | 1 | | | 1 |
| F - Construction | | | | | | | |
| Construction | 1 | | | 1 | 1 | 1 | 4 |
| G – Trade and repair of motor vehicles and domestic appliances | | | | | | | |
| Sale and repair of motor vehicles | | | 1 | 1 | 1 | 1 | 4 |
| Wholesale trade and sales agents | | | | | | 1 | 1 |
| Retail and repair of domestic appliances | | | | | 1 | | 1 |
| I - Transports and communication | | | | | | | |
| Land transport | 1 | | 1 | | 1 | | 3 |
| K – Real estate, letting and business services | | | | | | | |
| Research and development | 1 | 1 | | 2 | | | 4 |
| L – Public administration | | | | | | | |
| Public administration | | | 1 | 1 | | 1 | 3 |